\documentclass[preprint,12pt]{elsarticle}
\usepackage{graphicx}
\usepackage{amssymb}
\journal{Nuclear Physics A}
\begin{document}

\begin{frontmatter}

\title{Hyperons and massive neutron stars:\\
 the role of hyperon potentials               }

\author{S.~Weissenborn}
\ead{s.weissenborn@thphys.uni-heidelberg.de}

\author{D.~Chatterjee}
\ead{d.chatterjee@thphys.uni-heidelberg.de}

\author{J.~Schaffner-Bielich \corref{cor1}\fnref{label1}}
\ead{schaffner@thphys.uni-heidelberg.de}
 \fntext[label1]{Phone: +49 (6221) 54-9418, Fax: +49 (6221) 54-9333}

\cortext[cor1]{Corresponding author}

\address{Institut f{\"u}r Theoretische Physik, University of Heidelberg,\\
Philosophenweg 16, D-69120 Heidelberg, Germany}

\begin{abstract}
The constituents of cold dense matter are still far from being understood. 
However, neutron star observations such as the recently observed pulsar PSR J1614-2230 with 
a mass of $1.97\pm0.04$ M$_\odot$ help to considerably constrain the hadronic equation of state (EoS). 
We systematically investigate the influence of the hyperon potentials on the stiffness of the EoS. 
We find that they have but little influence on the maximum mass compared to the inclusion of an additional 
vector-meson mediating repulsive interaction amongst hyperons. 
The new mass limit can only be reached with this additional meson regardless of the hyperon potentials. 
Further, we investigate the impact of the nuclear compression modulus and the effective mass of the nucleon at saturation density on the high density regime of the EoS. 
We show that the maximum mass of purely nucleonic stars is very sensitive to the effective nucleon mass but only very little to the compression modulus.
\end{abstract}


\end{frontmatter}

\section{Introduction}
 The equation of state (EoS) in the vicinity of saturation density is pretty well understood. But beyond
saturation, the theories of dense matter present uncertainties. Neutron stars provide a fantastic astrophysical
environment for testing theories of cold and dense matter. In the core of neutron stars, densities could reach
values of several times $10^{15}$ g cm$^{-3}$. At such high densities, the Fermi energies of the constituent particles
could exceed the rest masses of heavier particles, and hence favour the appearance of such particles in the core.
Further, as the timescales associated with neutron stars are much greater than those associated with weak interactions,
violation of strangeness conservation due to weak reactions in the core would result in the appearance of
strangeness-containing particles such as hyperons. By producing new degrees of freedom, the appearance of strange 
particles is expected to result in a softer EoS of dense matter in the neutron star interior. The highest neutron star
mass that can be supported depends crucially on the EoS. Observations of pulsars in neutron star binaries
provide a precise measurement of neutron star masses through general relativistic effects. The best determined pulsar
mass ($1.4414\pm0.0002 M_{\odot}$) is the Hulse-Taylor pulsar, and masses of most other pulsars are found
to be clustered around this canonical value. Recently, several neutron stars with larger masses have been discovered.
Radio timing observations of three Post Keplerian parameters led to
the most precise measurement of the mass of a millisecond pulsar of $1.667\pm0.021 M_{\odot}$ \cite{Freire11}.
Shapiro delay measurements from radio 
timing observations of the binary millisecond pulsar PSR J1614-2230 indicated 
a mass of 1.97$\pm$0.04$M_{\odot}$ of the neutron star \cite{Demorest10}. 
This is the largest rather precisely observed pulsar mass so far and thus poses the tightest reliable lower bound on the maximum mass of neutron stars. 
Any theory of ultradense matter requires that the EoS produce a maximum mass at least as high as this measured value, i.e.,
models with $M^{max} \rm (theo)$ $ < M^{max} \rm(obs)$ would be ruled out. \\
\indent According to existing models of dense matter, the presence of hyperons leads to a considerable softening of the EoS, 
resulting in a reduction of the maximum mass of the neutron star. With hyperons, including only the hyperon-nucleon interaction, 
Brueckner-Hartree-Fock (BHF) calculations obtain maximum masses of the order 1.47$M_{\odot}$ \cite{Vidana00}.
 The inclusion of the hyperon-hyperon interaction leads to a further softening of the EoS and reduces the obtained masses to 
1.34$M_{\odot}$ \cite{SchulzeRijken}. On inclusion of three-body interactions, the maximum mass achieved in the BHF framework is 1.26$M_{\odot}$ \cite{Baldo98,Baldo00}. 
By employing a recently constructed hyperon-nucleon potential, the maximum masses of neutron stars with hyperons is computed to be well below 1.4$M_{\odot}$ \cite{Djapo}. 
Another approach is to adopt a relativistic mean field (RMF) Model \cite{Glen85,Knorren,Schaffner96}, which we employ for the following investigation. 
Several attempts have been made using the RMF model to explain neutron star masses higher than 2$M_{\odot}$, 
by artificially increasing the hyperon vector coupling away from their SU(6) values \cite{Huber,Hofmann,Rikovska}. 
Dexheimer et al. \cite{Dexheimer08} obtained large neutron star masses using a chirally motivated model including fourth-order
self-interaction terms of the vector mesons $\omega, \rho$ and $\phi$. Recently, Bednarek et al. \cite{Bednarek2011} achieved a stiffening of the EoS 
by invoking quartic vector-meson terms proportional to $\omega^4, \rho^4, \phi^4$ and cross terms
like $\phi^2 \omega^2, \phi^2 \rho^2$ and $\omega^2 \rho^2$. 
When the stiffening on inclusion of hyperons is not sufficient, a transition to the quark phase can be considered to produce large 
maximum masses \cite{Klahn2011}. Bonanno and Sedrakian \cite{Bonanno2011} also succeeded in obtaining a large neutron star mass 
with a hyperon and quark core using a stiff EoS and vector repulsion among quarks.  \\
\indent As the parameters in the RMF model
are fitted to the saturation properties of infinite nuclear matter, extrapolation to higher densities and asymmetry involve uncertainties. 
Three of these properties - the saturation density, the binding energy and the asymmetry 
energy are more precisely known than the remaining ones - the effective nucleon mass and the compression modulus of nuclear matter.  The
uncertainty in the dense matter EoS is basically related to the uncertainty in these two saturation properties.
In this paper, we determine which of the two influences most the high density part of the EoS that
determines the highest attained neutron star mass. 
The parameters associated with attractive interaction among hyperons and nucleons are fitted to the potential depths of hyperons 
in nuclear matter, known from hypernuclear experiments. 
We investigate in this paper how the uncertainty in hyperon
potential depths influences the stiffness of the EoS and hence the maximum mass. 
We want to address the question: under which conditions can hyperons be present in massive neutron stars? We achieve this through
a controlled parameter study within the RMF Model.
In our model, we succeed in generating stiff EoS 
by taking small values of effective nucleon mass $m^*$, in the range of those obtained from fits to nuclei rather than bulk nuclear matter, and by including the strange vector meson $\phi$. 
In a subsequent paper \cite{SimonDebbie2}, we will focus on the assumption of the underlying symmetries that govern the repulsive interactions among hyperons and nucleons.\\
\indent This paper is organized in the following way: In Sec. 2, we describe the model to calculate the 
EoS. The parameters of the model are listed in Sec. 3. The results of our calculations are
discussed in Sec. 4, and the summary and conclusions are given in Sec. 5.
\section{Theoretical Model}
\indent One of the possible approaches to describe neutron star matter is to adopt a RMF model 
subject to chemical equilibrium and charge neutrality. For our investigation of nucleons and hyperons in neutron 
star matter we will choose the full standard ($J^P=\frac{1}{2}^+$) baryon octet 
as well as electrons and muons. 
In this model, baryon-baryon interaction is mediated by the exchange
of scalar ($\sigma$), vector ($\omega$) and isovector ($\rho$) mesons. 
The Lagrangian density is given by \cite{Schaffner96}
\begin{eqnarray}
{\cal L} &=& \sum_B \bar\Psi_{B}\left(i\gamma_\mu{\partial^\mu} - m_B
+ g_{\sigma B} \sigma - g_{\omega B} \gamma_\mu \omega^\mu
- g_{\rho B}
\gamma_\mu{\mbox{\boldmath t}}_B \cdot
{\mbox{\boldmath $\rho$}}^\mu \right)\Psi_B\nonumber\\
&& + \frac{1}{2}\left( \partial_\mu \sigma\partial^\mu \sigma
- m_\sigma^2 \sigma^2\right) - U(\sigma) + U(\omega) \nonumber\\
&& -\frac{1}{4} \omega_{\mu\nu}\omega^{\mu\nu}
+\frac{1}{2}m_\omega^2 \omega_\mu \omega^\mu
- \frac{1}{4}{\mbox {\boldmath $\rho$}}_{\mu\nu} \cdot
{\mbox {\boldmath $\rho$}}^{\mu\nu}
+ \frac{1}{2}m_\rho^2 {\mbox {\boldmath $\rho$}}_\mu \cdot
{\mbox {\boldmath $\rho$}}^\mu .
\end{eqnarray}
The isospin multiplets for baryons B are
represented by the Dirac spinor $\Psi_B$ with vacuum baryon mass $m_B$, isospin operator ${\mbox {\boldmath t}}_B$, and $\omega_{\mu\nu}$ and 
$\rho_{\mu\nu}$ are field strength tensors. 
To reproduce the saturation properties of nuclear matter, the scalar self-interaction term
\begin{equation}
U(\sigma) = \frac{1}{3} b \sigma^3 + \frac{1}{4} c \sigma^4
\end{equation}
is introduced \cite{BogutaStoecker}. We also included an additional self-interaction term 
\begin{equation}
U(\omega) = \frac{1}{4} d (\omega_{\mu} \omega^{\mu})^2
\end{equation}
for the vector field as proposed
by Bodmer \cite{Bodmer}. Due to inclusion of this term, the vector field increases proportional
to $\rho^{1/3}$ for high densities, where $\rho$ is the baryon density, instead of the linear dependence in absence of this term. This results in a good agreement
with Brueckner-Hartree-Fock calculations.
We will denote the above model with the three exchange mesons as ``model $\sigma\omega\rho$''. 
The hyperon-hyperon interaction is usually incorporated through the exchange of additional strange scalar ($\sigma^*$) 
and strange vector ($\phi$) mesons, again the scalar meson being responsible for attractive and the vector meson for repulsive interactions respectively:
\begin{eqnarray}
{\cal L}_{YY} &=& \sum_B \bar\Psi_{B}\left(
g_{\sigma^* B} \sigma^* - g_{\phi B} \gamma_\mu \phi^\mu \right)\Psi_B\nonumber\\
&& + \frac{1}{2}\left( \partial_\mu \sigma^* \partial^\mu \sigma^*
- m_{\sigma^*}^2 {\sigma^*}^2\right) \nonumber\\
&& -\frac{1}{4} \phi_{\mu\nu}\phi^{\mu\nu}
+\frac{1}{2}m_\phi^2 \phi_\mu \phi^\mu.
\end{eqnarray}
Since it is our goal to obtain the stiffest possible EoS within the model, we will only make use of the $\phi$ but omit the $\sigma^*$. 
This choice is also in accordance with $\Lambda \Lambda$-hypernuclear data, where now it is clear that the
$\Lambda \Lambda$ interaction is only weakly attractive (see Ref \cite{Gal2011} for a discussion).
The model obtained with this additional Lagrangian we will call ``model $\sigma\omega\rho\phi$''. 
The calculation is performed using the mean field approximation \cite{Ser}. 
The effective baryon mass is given by 
$m_B^*=m_B - g_{\sigma B}\sigma$ while the chemical potential of baryon $B$ is
$\mu_{B} = (k^2_{F_{B}} + m_B^{* 2} )^{1/2} + g_{\omega B} \omega_0 + g_{\phi B} \phi_0 + t_{3B}g_{\rho B} \rho_{03}/2$, 
with $k_{F_{B}}$ denoting the corresponding Fermi momentum. 
Charge neutrality is described by the condition
\begin{equation}
Q = \sum_B q_B n_B -n_e -n_\mu =0~,
\end{equation}
where $n_B$ is the number density of baryon B, $q_B$ is the electric charge 
and $n_e$ and $ n_\mu$ are charge densities of electrons and muons respectively.
We consider the hadronic matter to behave like an ideal fluid. 
The energy-momentum tensor for such a fluid yields
the EoS, defined by the relationship between the total energy density  
\begin{eqnarray}
{\varepsilon}
&=& \frac{1}{2}m_\sigma^2 \sigma^2 
+ \frac{1}{3} b \sigma^3 + \frac{1}{4} c \sigma^4 + \frac{3}{4} d \omega_0^4 \nonumber \\ 
&& + \frac{1}{2} m_\omega^2 \omega_0^2 + \frac{1}{2} m_\phi^2 \phi_0^2 
+ \frac{1}{2} m_\rho^2 \rho_{03}^2 \nonumber\\
&& + \sum_B \frac{2J_B+1}{2\pi^2} 
\int_0^{k_{F_B}} (k^2+m^{* 2}_B)^{1/2} k^2 \ dk \nonumber\\
&& + \sum_{l=e^-,\mu^-} \frac{1}{\pi^2} \int_0^{K_{F_l}} (k^2+m^2_l)^{1/2} k^2 \ dk,
\end{eqnarray}
and the pressure
\begin{eqnarray}
P &=& - \frac{1}{2}m_\sigma^2 \sigma^2  
- \frac{1}{3} b \sigma^3 - \frac{1}{4} c \sigma^4 + \frac{1}{4} d \omega_0^4 \nonumber\\
&& + \frac{1}{2} m_\omega^2 \omega_0^2 + \frac{1}{2} m_\phi^2 \phi_0^2  
+ \frac{1}{2} m_\rho^2 \rho_{03}^2 \nonumber\\
&& + \frac{1}{3}\sum_B \frac{2J_B+1}{2\pi^2} 
\int_0^{k_{F_B}} \frac{k^4 \ dk}{(k^2+m^{* 2}_B)^{1/2}}\nonumber\\
&& + \frac{1}{3} \sum_{l=e^-,\mu^-} \frac{1}{\pi^2} 
\int_0^{K_{F_l}} \frac{k^4 \ dk}{(k^2+m^2_l)^{1/2}}~. 
\end {eqnarray}

\section{Parameters of the model}
\subsection{Nucleon-Meson coupling constants}
\indent There are five input parameters for the RMF models: the nuclear saturation density $\rho_0$ (or equivalently, the number density $n_0=\rho_0/m_N$) as well as the 
binding energy per baryon number B/A, the effective mass of the nucleon $m_N^*$, 
the nuclear compression modulus K and the asymmetry coefficient $a_{\it{sym}}$, all taken at saturation density. 
The parameters $g_{\sigma N}$, $g_{\omega N}$, $g_{\rho N}$, $b$ and $c$ 
of ``model $\sigma \omega \rho$'' are determined from the saturation properties of nuclear matter in an analytic way 
(see \cite{GlendenningBook} for explicit formulae). 
The binding energy, the asymmetry coefficient and the saturation density are well determined.
We set them to $n_0=0.153\:fm^{-3}$, $B/A=-16.3$ MeV, $a_{\it{sym}}=32.5$ MeV \cite{GM1}. 
The effective nucleon mass and the compression modulus are less well known. 
{In principle, the parameters of the model can be fitted to properties of nuclei, which result in low effective masses \cite{NL-Z}.
However, here we want to explore the full parameter range of the model, as model parameters fitted to bulk nuclear matter
and usually adopted in the community have large effective mass parameters.
Hence for this study, we use the parameter sets such as GM1 ($K$=300 MeV, $m_N^*/m_N=0.7$) and GM3 ($K$=240 MeV, $m_N^*/m_N=0.78$) 
which are fitted to the properties of bulk nuclear matter and have rather large effective nucleon masses in the range $m_N^*/m_N\approx0.7-0.8$ \cite{GM1}, and also
the parameter sets fitted to the properties  of nuclei, e.g. the 
whole NL parameter family and TM1 which yield very low values of $m_N^*/m_N\approx0.5-0.65$ \cite{TM1,NL3}.

The value of compression modulus for symmetric nuclear matter can be extracted from the energies of the isoscalar giant monopole
resonances (ISGMR) in nuclei (see e.g. \cite{compmodvalues} and references therein). Systematic studies of ISGMR energies
in various nuclei suggest a value of $231 \pm 14$ MeV for the incompressibility of symmetric nuclear matter \cite{Shlomo, Lui}. 
However, there are strong surface effects, as well as large uncertainties in the value of the compression modulus
for higher densities and asymmetric nuclear matter. Another semiempirical parameter of nuclear matter crucial for
determining the stiffness of the EoS is the density dependence of symmetry energy, denoted by $L$. We computed the values of the 
density dependence of symmetry energy $L$ for the sets discussed above, and they were
found to be fairly comparable with the values existing in the literature \cite{ChenKoLi,Centelles2010}. These however
differ from the values of $L$ extracted from isospin diffusion data ($L = 88 \pm 25$ MeV) \cite{Lisodiff} and isoscaling data 
($L \approx 65$ MeV) \cite{Lisoscalar}. We note here that when the effective mass $m^*/m$ is varied between 0.55 and 0.8, 
$L$ changes from 108 to 94 MeV ($\pm$ 1 MeV depending on the
compressibility).  

\subsection{Hyperon-Meson coupling constants}
The hyperon coupling constants $g_{\omega Y}$,  $g_{\rho Y}$ and $g_{\phi Y}$ are determined by using SU(6) symmetry (the
quark model) \cite{Dover85,Schaffner94}:
\begin{eqnarray}
\frac{1}{3} g_{\omega N} = \frac{1}{2} g_{\omega \Lambda} = \frac{1}{2} g_{\omega \Sigma} &=& g_{\omega \Xi}\:, \nonumber \\
g_{\rho N} = \frac{1}{2} g_{\rho \Sigma} &=& g_{\rho \Xi}\:, \nonumber \\
g_{\rho \Lambda} &=& 0\:, \nonumber \\
2 g_{\phi \Lambda} = 2 g_{\phi \Sigma} = g_{\phi \Xi} &=& - \frac{2 \sqrt{2}}{3} g_{\omega N}~.
\end{eqnarray}
 The scalar meson-hyperon coupling constants $g_{\sigma Y}$ are adjusted to the potential 
depths $U_Y^{(N)}$ felt by a hyperon Y in a bath of nucleons at saturation \cite{GM1} 
following the relation
\begin{equation}
 U^{(N)}_Y = g_{\sigma Y} \sigma^{eq} + g_{\omega Y} \omega_0^{eq}~,
\end{equation}
where the hypernuclear potential depths in nuclear matter $U^{(N)}_Y$ are fixed in accordance with the available hypernuclear data.
The best known hyperon potential is that of the $\Lambda$, having a value of about 
$U_\Lambda^{(N)}=-30$ MeV \cite{Millener88,Mares,Schaffner92}.
In case of the $\Sigma$s and the $\Xi$s the potential depths are not as firmly known as for the $\Lambda$. 
The following values $U_\Sigma^{(N)}=+30$ MeV, $U_\Xi^{(N)}=-18$ MeV are generally adopted from the hypernuclear experimental data (see \cite{SchafGal} for a discussion).

\section{Results and Discussions}
\subsection{Hyperon Potential Depths}\label{potentialdepths}
\indent We begin our investigations with ``model $\sigma \omega \rho$'', namely with a $\sigma-\omega$ model including 
non-linear scalar and vector self-interactions as well as $\rho$ mesons and the full baryon octet. In the previous section, we already mentioned
that there are uncertainties in the values for the potential depths of hyperons in nuclear matter, obtained from hypernuclear
experiments. We vary the potentials $U_\Sigma^{(N)}$ and $U_\Xi^{(N)}$ systematically to study how they influence the stiffness of the hadronic EoS 
and the limiting neutron star mass. 
\subsubsection{Varying $U_\Sigma^{(N)}$}
\indent We start with a fixed $\Xi$ potential depth of $U_\Xi^{(N)}=-28$ MeV \cite{Schaffner96}. 
The potential depth of $\Sigma$ is varied in the range $-40$ MeV $\leq U_\Sigma^{(N)}\leq +40$ MeV. 
We plot in Fig. \ref{figuresigmafreeeos} the EoS for different values of $U_\Sigma^{(N)}$ (lower branch) for the 
GM1 parameter set and 
also for the same model with additional $\phi$ meson (upper branch).  
For the sake of clarity only the cases $U_\Sigma^{(N)}=-40,-30,-20,0,+40$ MeV are displayed.
\begin{figure}
\includegraphics[width=12 cm]{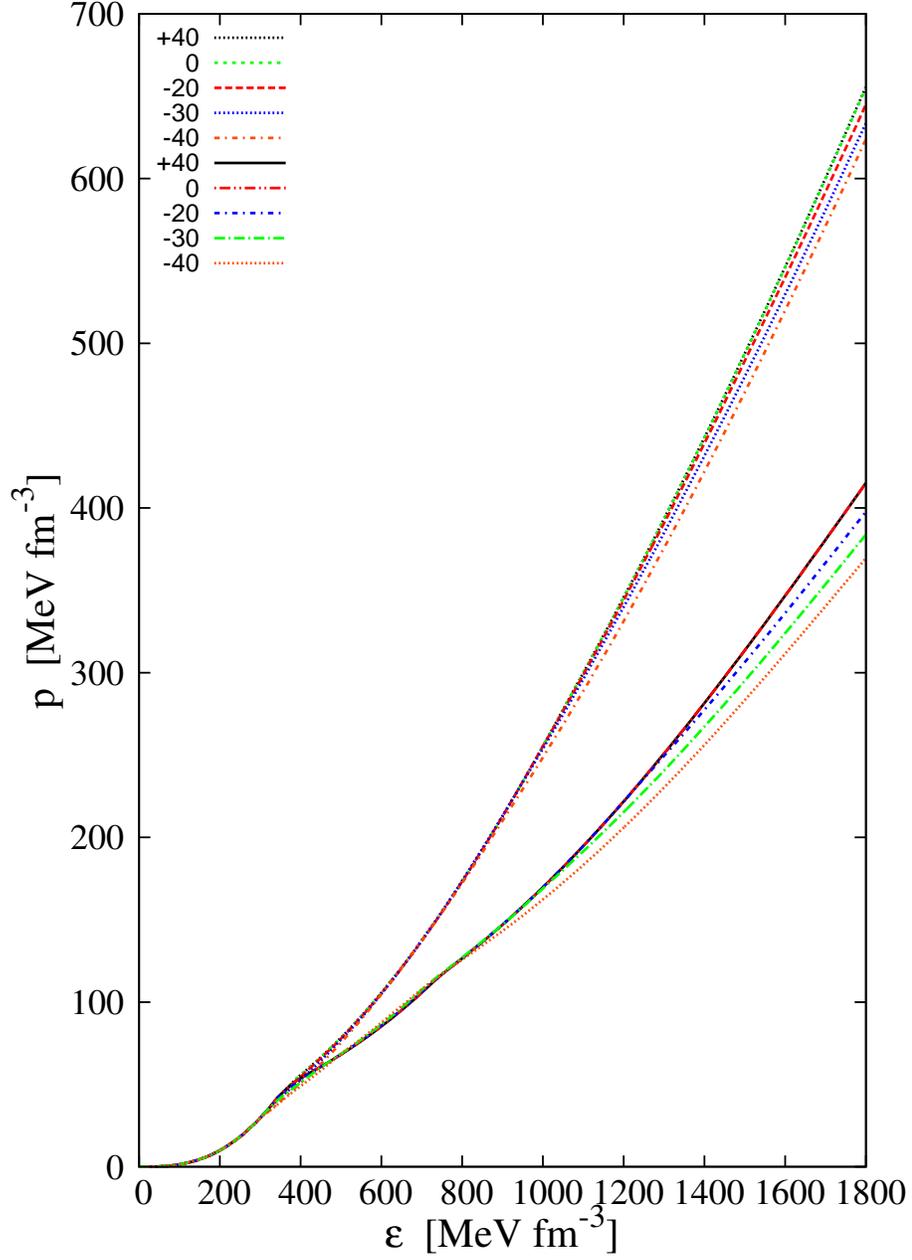}
\caption[EoS for different potential depths $U_\Sigma^{(N)}$.]
{EoS for different potential depths $U_\Sigma^{(N)}$ in MeV, indicated by the corresponding numbers in the key. 
The lower branch is obtained for ``model $\sigma \omega \rho$'' for the GM1 parameter set while the upper branch is obtained including additionally the $\phi$ meson.}
\label{figuresigmafreeeos}
\end{figure}

\indent It can be inferred from the figure that for less deep potentials $U_\Sigma^{(N)}$ the EoS 
becomes stiffer for each model. However, this is only true for negative potentials, since the values 
$U_\Sigma^{(N)}=0$ MeV and $U_\Sigma^{(N)}=+40$ MeV basically result in the same EoS. 
For negative values of the potential $U_\Sigma^{(N)}$, the $\Sigma$s 
are bound in nuclear matter and the deeper the potential is, the more attractive must be the 
effective mesonic interaction and thus the softer the EoS. 
For $U_\Sigma^{(N)}\geq 0$ MeV, the $\Sigma$s are no longer bound to nuclear matter and the 
effective mesonic interaction becomes more and more repulsive for increasing $U_\Sigma^{(N)}$. 
This would in principle stiffen the EoS. 
However, up to neutron star densities, i.e. about $n_B\lesssim (4-7)n_0$, the $\Sigma$s are not present in hadronic matter if the 
potential is repulsive and thus the EoS up to these densities becomes insensitive to the actual value of $U_\Sigma^{(N)}$. 
The composition of hadronic matter for positive and negative values of $\Sigma$ potential depth has already been investigated
in \cite{Knorren,Schaffner96,Schaffner08}.\\
\begin{figure}
\includegraphics[width=12cm]{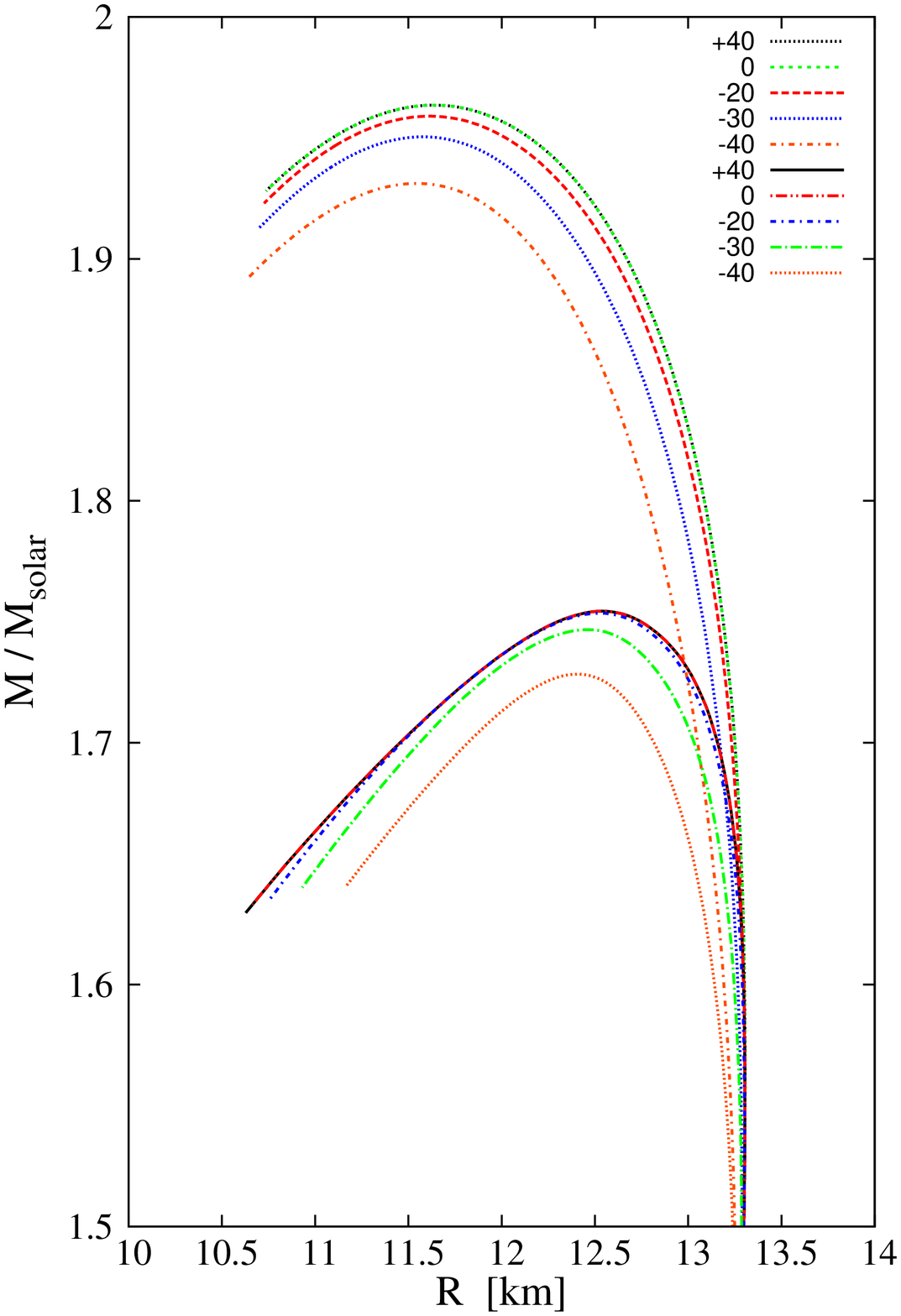} 
\caption[Mass radius relations of neutron stars for different potential depths $U_\Sigma^{(N)}$]
{Mass radius relations for neutron stars obtained with the EoS from Fig. \ref{figuresigmafreeeos}. 
The variation of $U_{\Sigma}^{(N)}$ in ``model $\sigma \omega \rho$`` cannot account for the observed neutron star mass limit (lower branch), unless 
the $\phi$ meson is included in the model (upper branch).}
\label{figuresigmafreemr}
\end{figure}
\indent The repulsive potentials $U_\Sigma^{(N)}\geq0$ give 
the highest maximum star masses for each parameter set (also for others which we did not plot, as e.g. TM1 and NL3). 
In Fig. \ref{figuresigmafreemr} we plot the parts of the mass-radius relations around the 
maximum mass star configurations for the EoS in Fig. \ref{figuresigmafreeeos}. 
The maximum mass for the lower branch of curves in Fig. \ref{figuresigmafreemr} corresponding to ``model $\sigma \omega \rho$''
is ${M}_{\it{max}}=1.75 {M}_\odot$.
We note, that the variation of $U_\Sigma^{(N)}$ in ``model $\sigma \omega \rho$'' for GM1 does not help 
to increase the maximum neutron star mass above the limiting value of 1.97$\pm$0.04${M}_{\odot}$. 
Investigating other parameter sets like TM1 would not solve the problem since the maximum masses are comparable to those of GM1. 
The various NL-models would be good candidates, since they are amongst the stiffest available EoS \cite{Schaffner96}. 
However, the NL-models have (like all models which are fitted to finite nuclei) 
very small effective nucleon masses already at saturation density ($m^*/m\approx 0.55-0.65$). 
At around five times nuclear saturation density the $\sigma-$fields are so large that the effective nucleon mass becomes zero or in principle even negative. 
This causes a well-known instability which makes it impossible to find a physical solution to the RMF equations \cite{Schaffner96}.
However, the critical density for the onset of the instability is higher than the maximum central
density in the neutron star interior, and hence this is not a relevant problem for the present investigation. \\
\indent The ``model $\sigma \omega \rho \phi$'' gives us the upper bundle of EoS in Fig. \ref{figuresigmafreeeos}. 
In Fig. \ref{figuresigmafreemr} the upper branch of the mass-radius relations corresponds to these EoS from ``model $\sigma \omega \rho \phi$''. 
While the variation of $\Sigma$ potential only varies the maximum mass in the range $\Delta_M \approx$ 0.03 ${M}_{\odot}$, 
the impact of $\phi$ on the maximum mass is an order of magnitude larger: 
the maximum mass is increased by nearly 0.21 ${M}_\odot$ to ${M}_{\it{max}}$=1.96 ${M}_\odot$. 
 We should also mention that for purely nucleonic matter EoS the obtained maximum masses are even higher 
 (about 2.4${M}_{\odot}$ for GM1 and nearly 2.8${M}_\odot$ for NL3). 
\subsubsection{Varying $U_\Xi^{(N)}$}
\indent We have seen that the potential depth $U_\Sigma^{(N)}$ has but a little influence on the maximum mass of a hadronic neutron star. 
As it does not play an important role in our investigations,
we shall fix  it to the value $U_\Sigma^{(N)}$=+30 MeV \cite{Schaffner96} 
which means that the $\Sigma$s are not present and the EoS is as stiff as possible. 
We then vary the $\Xi$ potential $U_\Xi^{(N)}$ in the range $-40{MeV}\leq U_\Xi^{(N)}\leq$+40MeV. 
As before, we compute the EoS for the GM1 parameter set for ``model $\sigma \omega \rho$'' and ``model $\sigma \omega \rho \phi$'' for different $\Xi$ potentials. 
We plot in Fig. \ref{figurexifreeeos} the results for the values of $U_\Xi^{(N)}$ from -40 MeV to +40 MeV in steps of 20 MeV. 
\begin{figure}
 \includegraphics[width=12cm]{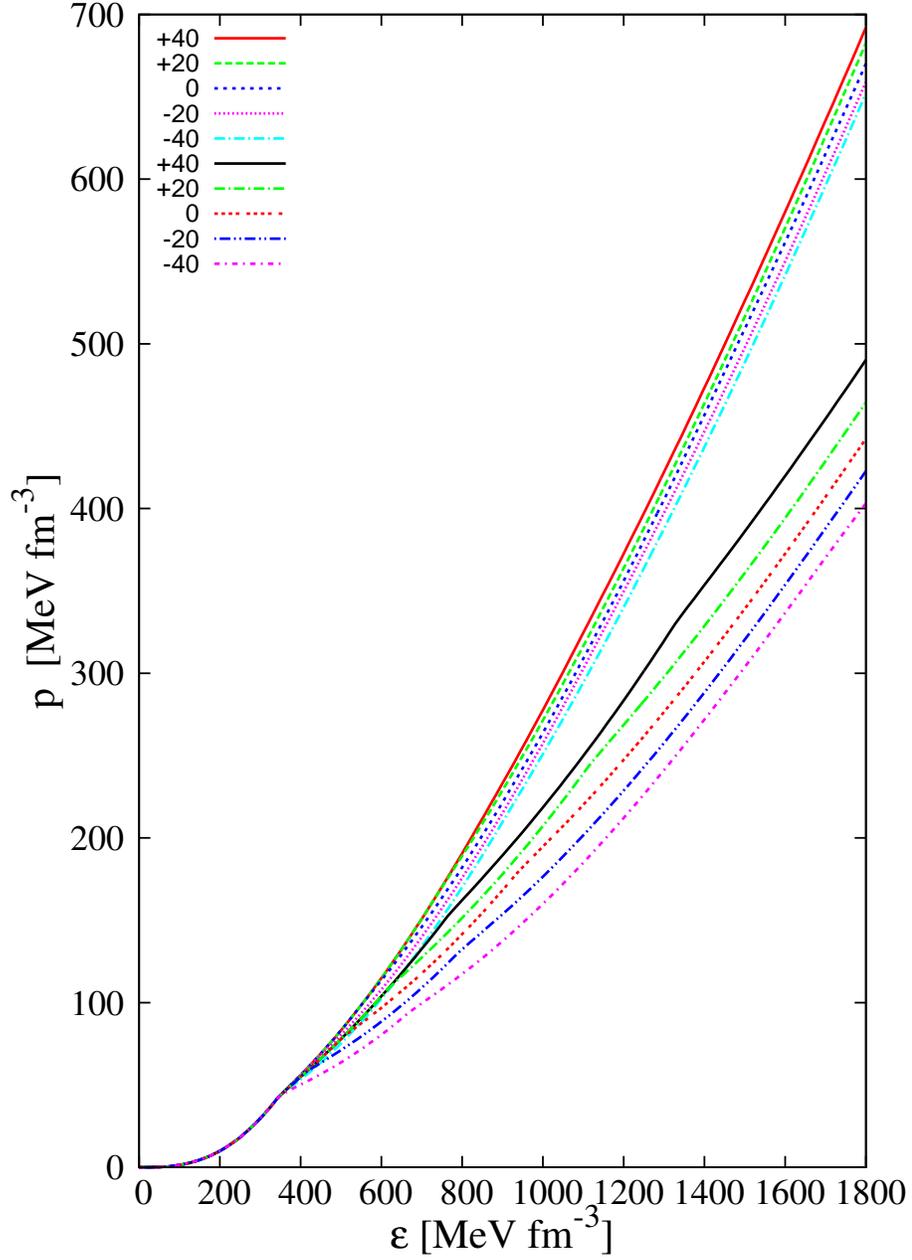}
\caption[Equations of state for different potential depths $U_\Xi^{(N)}$]
{EoS for different potential depths $U_\Xi^{(N)}$. The lower bundle of curves corresponds to ``model $\sigma \omega \rho$'' and the upper to ``model $\sigma \omega \rho \phi$''. 
All EoS are obtained for GM1 parameter set.}
\label{figurexifreeeos}
\end{figure}
\begin{figure}
 \includegraphics[width=12cm]{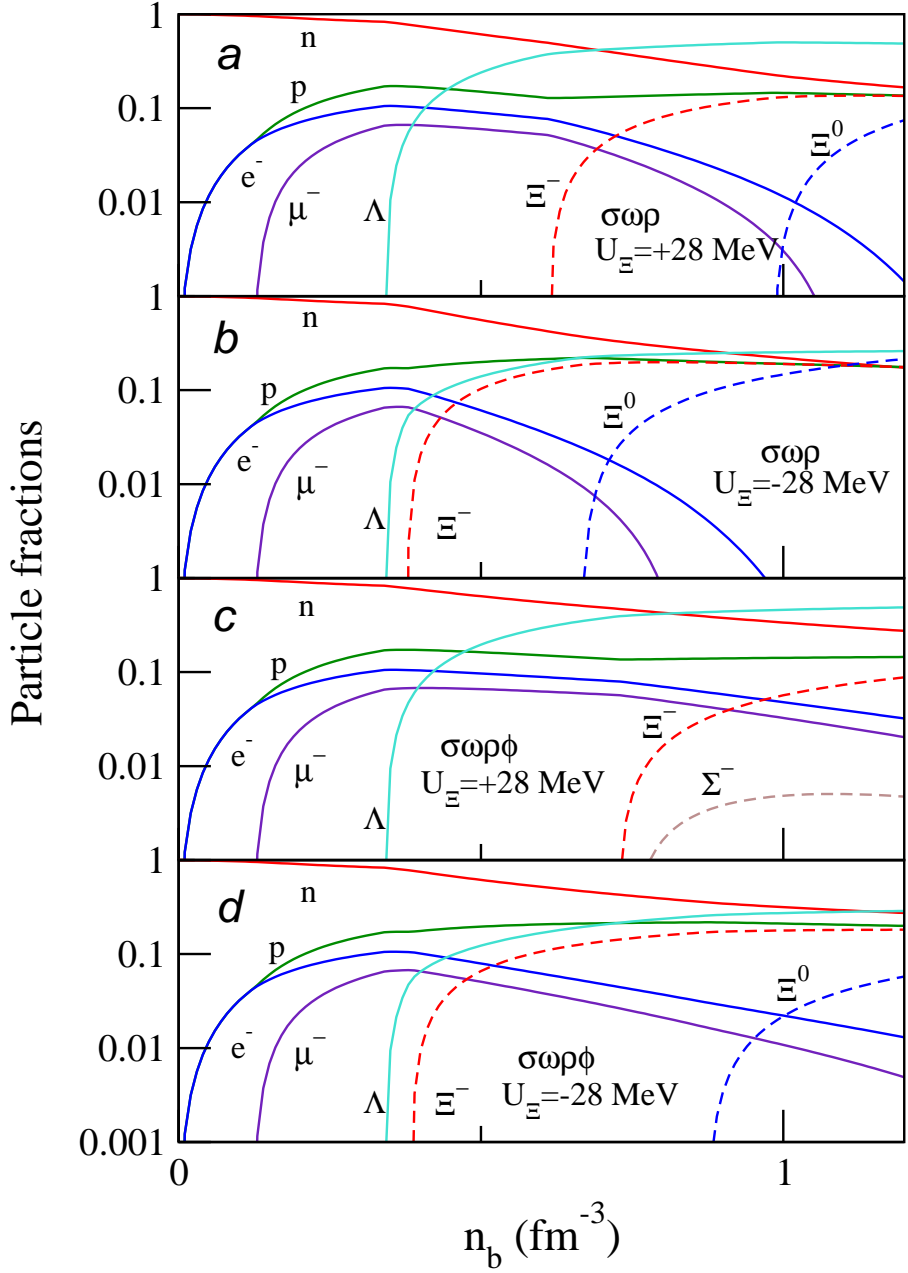}
\caption[Particle fractions for different potential depths $U_\Xi^{(N)}$]
{Particle fractions for ``model $\sigma \omega \rho$'' for potential depths $U_\Xi^{(N)}$ = +28 MeV (a) and -28 MeV (b) are displayed.
Similarly, particle fractions for ``model $\sigma \omega \rho \phi$'' are shown for $\Xi$ potentials +28 MeV (c) and -28 MeV (d).
The threshold for appearance of the $\Xi$ hyperons is pushed to higher densities with increasing 
$\Xi$ potential.}
\label{figureuxifrac}
\end{figure}

\indent We see that for increasing $\Xi$ potential, the EoS stiffens. 
A difference to the influence of the $\Sigma$ potential depth is found for positive values of $U_\Xi^{(N)}$: 
here the EoS remains sensitive to the value of the potential
due to the contributions from $\Xi^-$ at neutron star densities while for even higher densities both $\Xi$s contribute.
In Fig. \ref{figureuxifrac}, we display how the composition of the neutron star core varies for various values of $\Xi$ potential
in models ``$\sigma \omega \rho$'' and ``$\sigma \omega \rho \phi$''. It is clear from the figure that for positive $\Xi$ potential,
$\Lambda$ hyperons dominate particle fractions at high densities, while for negative $U^{(N)}_{\Xi}$ the particle fractions from 
all hyperons contribute. As $\Xi$ potential increases from negative to positive values, the threshold density for appearance
of $\Xi^-$ and $\Xi^0$ hyperons is pushed to higher densities, and this effect further increases on inclusion of repulsion
from the $\phi$ mesons.
\\
\begin{figure}
 \includegraphics[width=12cm]{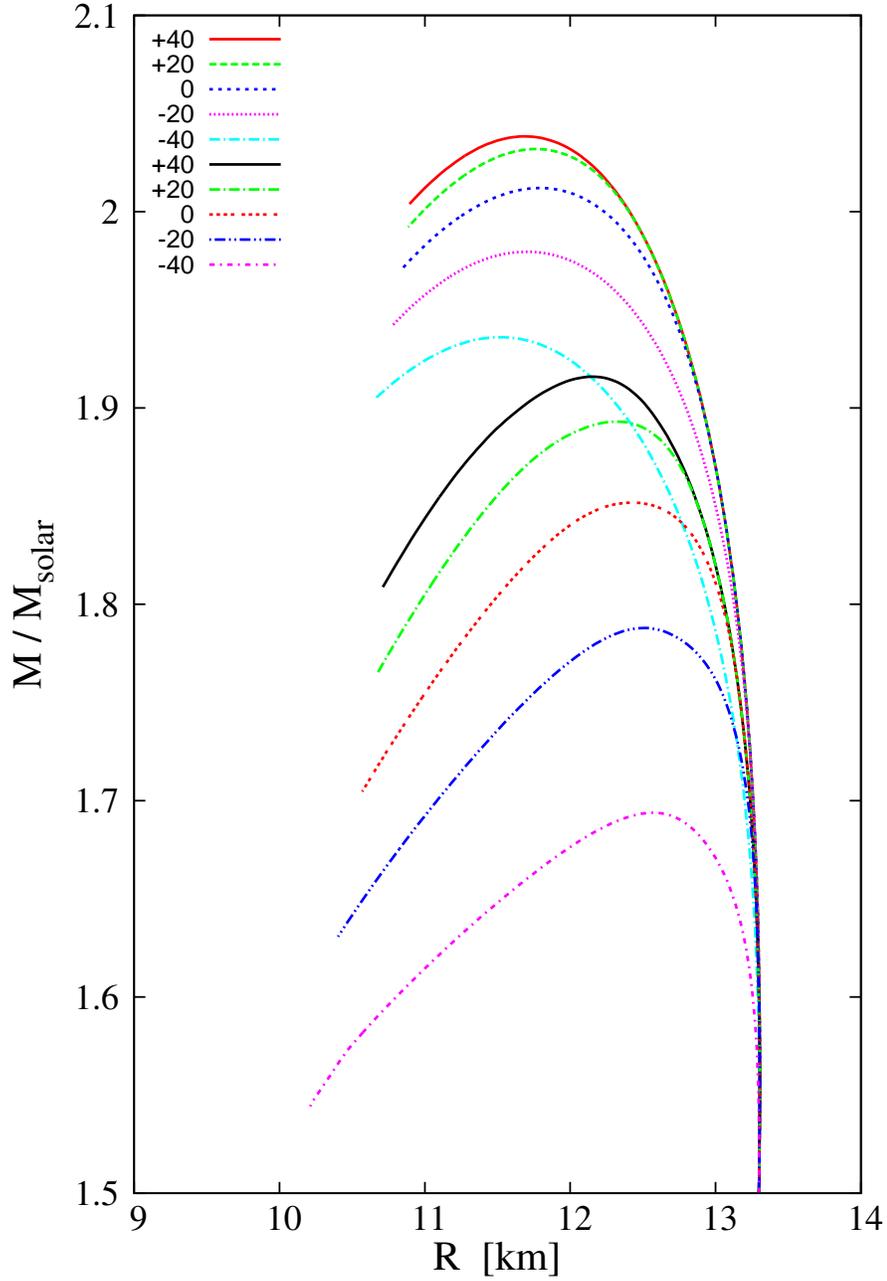}
\caption[Mass-radius relations for different values of $U_\Xi^{(N)}$]
{Mass-radius relations obtained from the EoS of Fig. \ref{figurexifreeeos}. 
As before, the variation of $U_{\Xi}^{(N)}$ in ``model $\sigma \omega \rho$`` cannot explain the observed neutron star mass limit (lower branch), unless 
the $\phi$ meson is included (upper branch).}
\label{figurexifreemr}
\end{figure}
\indent  In Fig. \ref{figurexifreemr} we plot the 
mass-radius relations of the neutron stars obtained from the various EoS in Fig. \ref{figurexifreeeos}. 
Within ``model $\sigma \omega \rho$'' the maximum masses range from 1.69 ${M}_{\odot}$ for $U_\Xi^{(N)}=-40$ MeV 
to 1.92 ${M}_{\odot}$ for $U_\Xi^{(N)}$=+40 MeV, i.e. over an interval of $\Delta_M\approx$0.22 ${M}_{\odot}$. 
The maximum mass for the case $U_\Xi^{(N)}$=+40 MeV including $\phi$ is 2.04${M}_{\odot}$.
Although the value of the $\Xi$ potential depth helps to stiffen the EoS far better than the $U_\Sigma^{(N)}$ 
it is still not enough to be consistent with the mass of PSR J1614-2230 for ``model $\sigma \omega \rho$''. 
However, the mass constraint can be fulfilled within ``model $\sigma \omega \rho\phi$'' for the GM1 parameter set. 
\subsubsection{The  $U_\Sigma^{(N)}-U_\Xi^{(N)}$ Plane}
\indent We combine the variations of $U_\Sigma^{(N)}$ and  $U_\Xi^{(N)}$ and plot in Fig. \ref{figurepotentialcontour} 
lines of constant maximum mass in the $U_\Sigma^{(N)}-U_\Xi^{(N)}$ plane within the range $-40$ to $+40$ MeV for both potentials. 
As before, we compare the results from ``model $\sigma \omega \rho$'' with those of ``model $\sigma\omega\rho\phi$''.
\begin{figure}
\vspace {-4cm}
\includegraphics[height=12cm,width=12cm,angle=270]{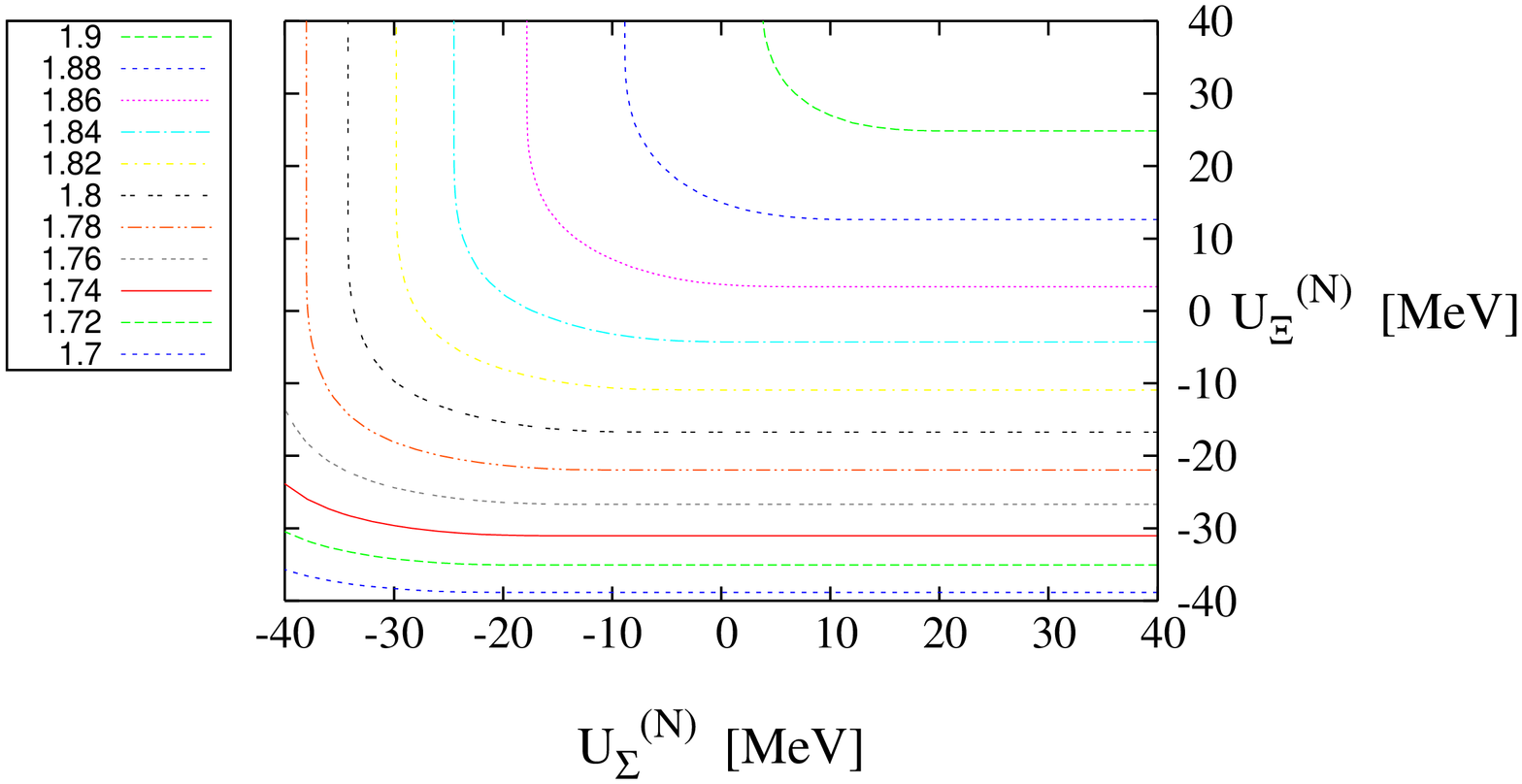} 
\vskip - 3 cm
\includegraphics[height=12cm,width=12cm,angle=270]{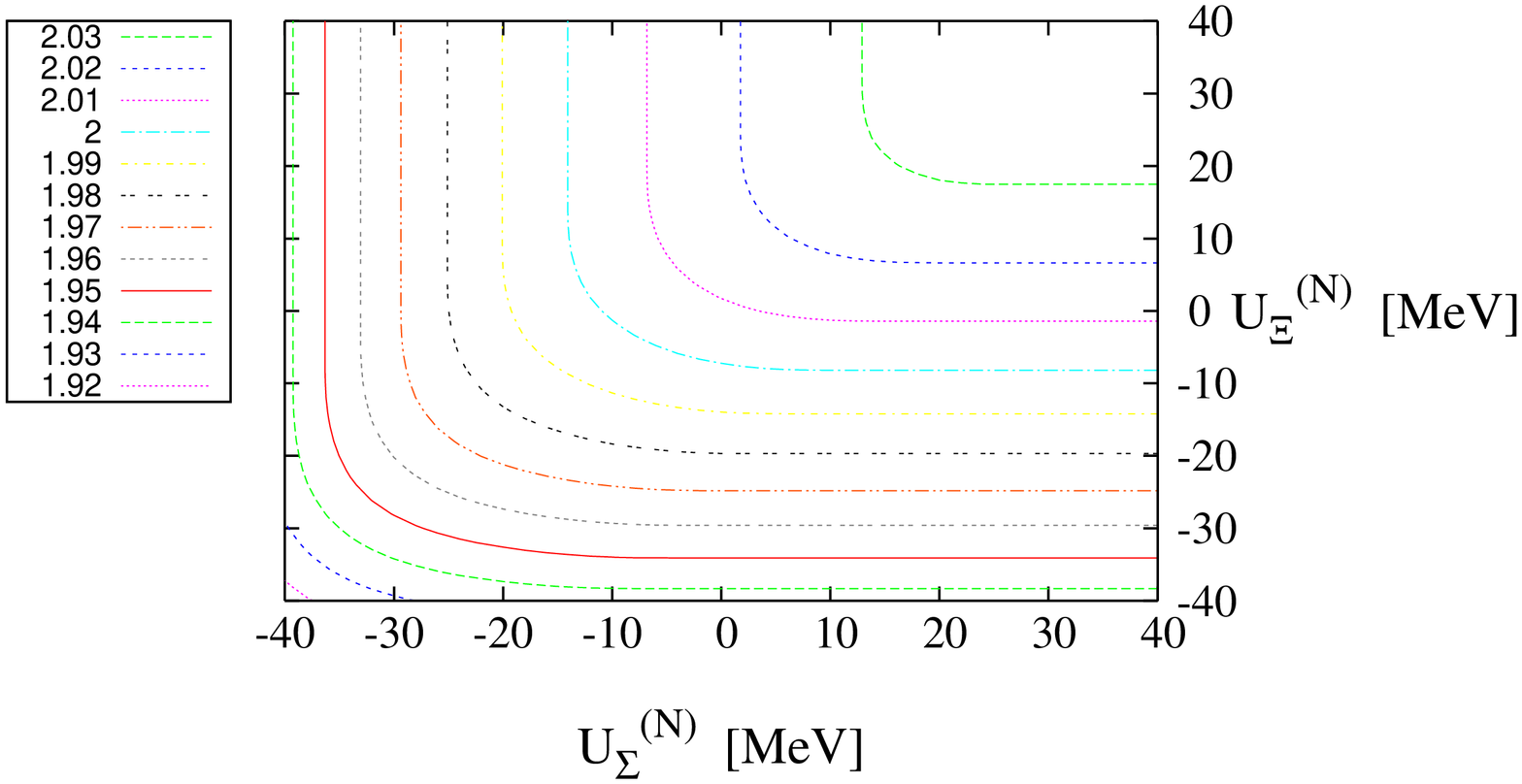}
\vskip - 1 cm
\caption[Contour plot: lines of constant maximum mass in the $U_\Sigma^{(N)}-U_\Xi^{(N)}$ plane.]
{Lines of constant maximum mass (key numbers in M$_\odot$) in the $U_\Sigma^{(N)}-U_\Xi^{(N)}$ plane. 
Upper plot for ``model $\sigma \omega \rho$'' in GM1 parameterization, lower one for the same model including also $\phi$ mesons.}
\label{figurepotentialcontour}
\end{figure}
In the light of our previous discussions we can quickly explain the shape of the lines in the plot: 
On each side of the $U_\Sigma^{(N)}=U_\Xi^{(N)}$ line the maximum masses are dominated by 
that potential which is smaller and are rather insensitive to the value of the larger potential. 
This means, that in the region where $U_\Sigma^{(N)}>U_\Xi^{(N)}$ the lines of constant 
maximum mass are parallel to the $U_\Sigma^{(N)}-$axis. 
In the other part of the plot, where $U_\Sigma^{(N)}<U_\Xi^{(N)}$, the behaviour is mirrored. 
Since the influence on the maximum masses is larger for smaller potentials, 
the mass lines (which are plotted for equidistant mass steps of 0.01${M}_{\odot}$) 
are densest towards the lower and the left parts of the plots but thin out with increasing potentials.
\indent We conclude this section by noting that for the whole range $-40$ MeV $\leq U_\Sigma^{(N)}$,$U_\Xi^{(N)}\leq +40$ MeV 
the maximum masses of ``model $\sigma \omega \rho$'' for the GM1 parameter set are not compatible with the observed pulsar mass $1.97\pm$0.04${M}_{\odot}$. 
Inclusion of the $\phi$ meson helps to sufficiently stiffen the EoS and the whole range of $-40{MeV}\leq U_\Sigma^{(N)}$,$U_\Xi^{(N)}\leq$+40MeV is now allowed. 
However, in this case the new mass limit does not help to restrict the parameter range of the hyperon potentials.
\subsection{Compression Modulus and Effective Mass}
\begin{figure}
\includegraphics[width=13cm]{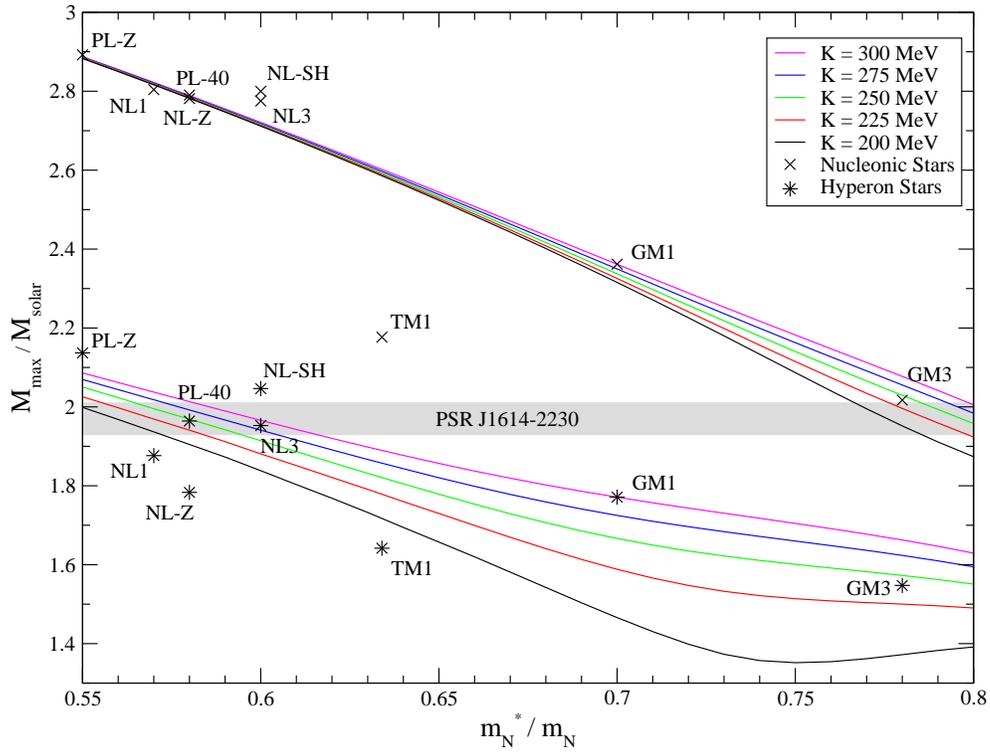}
\caption[$M_{\it{max}}$ vs. $m_N^*$ for different values of K for nuclear matter stars]
{Maximum masses of neutron stars as functions of the effective nucleon mass for different compression moduli using ``model $\sigma\omega\rho$``. 
The upper curves refer to purely nucleonic stars while for the lower curves hyperons are included.}
\label{figuremefffree}
\end{figure}
\indent In the previous section, we found that the variation of hyperon potentials were not sufficient to explain
the observed large neutron star mass. We then proceed to find out how the maximum mass depends on the nucleon coupling constants. We vary the 
least well known of the saturation parameters, namely the effective nucleon mass $m_N^*$ and the compression modulus K. 
We first consider purely nucleonic stars, and then repeat the exercise including hyperons
in the ``model $\sigma\omega\rho$''. 
In Fig. \ref{figuremefffree} we plot the maximum masses as a function of the effective nucleon mass at saturation $m_N^*/m_N$ for several values of K. 
We see that the maximum masses for the nucleonic case (upper bunch of curves) show a strong dependence on the effective nucleon mass: 
they drop from 2.88${M}_{\odot}$ at $m_N^*/m_N=0.55$ to 
1.87-2.01${M}_{\odot}$ at $m_N^*/m_N=0.8$. 
From Fig. \ref{figuremefffree} it is clear that for moderate effective masses, $m_N^*/m_N\approx0.6-0.7$, 
the compression modulus gains a little influence on the maximum mass for pure nucleonic stars, 
and causes a mass difference of less than 0.05${M}_{\odot}$.
Above this value of the effective mass, the compression modulus causes a larger splitting of the maximum masses
until at $m_N^*/m_N=0.8$ the mass difference is about 0.1${M}_{\odot}$.
For the hyperonic case (lower bunch of curves), the maximum masses range from 2.0-2.1${M}_{\odot}$ 
at $m_N^*/m_N=0.55$ to about 1.39-1.63${M}_{\odot}$ at $m_N^*/m_N=0.8$.
The mass splitting at $m_N^*/m_N\approx0.65$ for different K is about 0.2${M}_{\odot}$, increases to nearly 0.35${M}_{\odot}$
at $m_N^*/m_N\approx0.75$, and then again decreases to about 0.25${M}_{\odot}$ at $m_N^*/m_N\approx0.8$.
For comparison we also mark in Fig. \ref{figuremefffree} the maximum masses and the effective 
masses of several other RMF sets fitted to properties of nuclei, like TM1, NL3 or NL-SH \cite{NL3,Sugahara94,NL-Z,NL-SH,PL-Z}. 
For the ``model $\sigma\omega\rho\phi$'', the maximum masses range from 2.4 $M_{\odot}$ at $m_N^*/m_N\approx0.55$ to
1.6 $M_{\odot}$ at $m_N^*/m_N\approx0.8$. We also note that the results do not change
substantially from the results obtained using ``model $\sigma\omega\rho$'' when we include the $\sigma^*$ meson (in ``model $\sigma\omega\rho\sigma^*\phi$'').
Contrary to the statement made in \cite{GlendenningBook,Glendenning86}, that K as well as $m_N^*$ 
''strongly influence the high-density behaviour of the equation of state of symmetric nuclear matter'' this being an  
``effect [which] shows up in neutron rich and neutron star matter``, we find that the compression modulus 
has nearly no influence on the high-density behaviour of the nucleonic EoS
whereas the effective mass has a stark impact. This finding is well known for the RMF model \cite{BogutaStoecker,Waldhauser88}. 
The effect of the compressibility on hyperonic stars is however remarkable, although this dependence also varies
considerably with the effective nucleon mass. In \cite{GlendenningBook,Glendenning86}, the investigated range of the effective
nucleon mass was 0.7-0.8 for the hyperonic case, in which K has a greater influence on the maximum mass than  $m_N^*/m_N$,
which we also confirm in our study. However, we also extend our investigation for a broader range of $m_N^*/m_N$, and find that
this conclusion no longer holds for low effective nucleon masses. 
Also, at large $m_N^*/m_N$ where K is influential the maximum masses of hyperonic stars 
do not comply with the limiting pulsar mass of 1.97$\pm$0.04${M}_{\odot}$. 
In ``model $\sigma\omega\rho$'' with hyperons this mass is only reached for low values of the effective nucleon mass 
$m_N^*/m_N \leq 0.6$.
Within the  ``model $\sigma\omega\rho \phi$'' the mass constraint can only be achieved for $m_N^*/m_N \leq 0.7$. 
 In the case of nucleonic stars, for $m_N^*/m_N \gtrapprox 0.78$ the pulsar mass constraint
cannot be reached if the compression modulus is below 225 MeV.  
\\
\begin{figure}
 \includegraphics[height=12cm,width=12cm,angle=270]{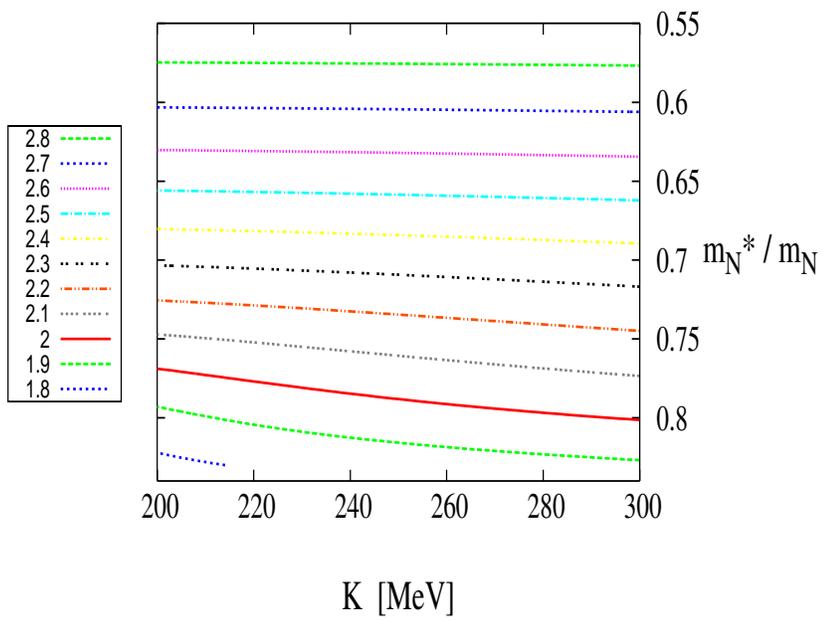}
\caption[Lines of constant maximum mass in the $m_N^*-K$ plane.]
{Lines of constant maximum mass of purely nuclear matter stars in the $m_N^*-K$ plane.}
\label{figuremeffKcontour}
\end{figure}
\indent To explore the whole range of K continuously, we plot lines of constant maximum mass of nucleonic stars in the $m_N^*-K$ plane in Fig. \ref{figuremeffKcontour}. 
The results support our previous observation: the lines of constant maximum mass are nearly parallel to the x-axis, which means that the 
compression modulus does not influence the stiffness of the EoS at neutron star densities. 
Only for effective masses above $m_N^*/m_N\geq0.7$ do the lines become slightly slanted and show that a 
higher compression modulus then slightly stiffens the EoS. 
We note, that in the investigated case of purely nucleonic neutron stars 
the observed mass of PSR J1614-2230 requires an effective nucleon mass at saturation density of maximally $m_N^*/m_N\leq0.82$ at $K=300$ MeV 
down to  $m_N^*/m_N\leq0.78$ at $K=200$ MeV.
\section{Summary} 
\indent The possibility of existence of hyperons in massive neutron stars has been investigated in this paper. 
Within hadronic models, we varied the coupling strengths of the scalar $\sigma$ meson to the $\Sigma$ and $\Xi$ hyperons by 
changing the potential depths of these hyperons in nuclear matter. 
Meanwhile, we assumed SU(6) relations for the vector couplings and kept the $\Lambda$ potential fixed. 
We found that for deeper potential depths the EoS becomes softer and the maximum masses correspondingly smaller.  
Which of the two potentials has a larger impact on the stiffness of the EoS depends mainly on their relative values:
if $U_{\Xi}^{(N)} < U_{\Sigma}^{(N)}$ the $\Xi$s will appear earlier (i.e. at lower densities) than the $\Sigma$s and therefore 
have the greater influence on the EoS. 
However, this cannot exactly be mirrored, since for $U_{\Xi}^{(N)} > U_{\Sigma}^{(N)}$ the $\Sigma^-$ might appear 
earliest of all $\Sigma$s and $\Xi$s, but as soon as the $\Xi^-$ appears, the number density of the $\Xi^-$ overtakes that of the $\Sigma^-$ by far. 
The overall effect of varying the two potential depths is rather small ($\Delta_M\approx$0.2${M}_{\odot}$ over the probed range).\\
\indent Since the impact of the potential depths on the EoS is not enough to raise the maximum masses of neutron stars for ``model $\sigma \omega \rho$'' for GM1 above the 
limit of $1.97\pm$0.04${M}_{\odot}$, we decided to increase the repulsive interaction amongst hyperons by including the vector $\phi$ meson in the RMF model. 
The additional repulsion between hyperons results in a sufficiently stiff EoS accompanied by an increase in the maximum masses of about $\Delta_M\approx$0.2${M}_{\odot}$.  
For the GM1 parameter set, the investigated area of the $U_{\Sigma}^{(N)}-U_{\Xi}^{(N)}$ plane complies now to almost full extent with the mass of PSR J1614-2230. \\
\indent In the case of purely nucleonic stars we varied the effective mass of the nucleon as well as the compression modulus at nuclear saturation density which are input parameters 
for the RMF model. 
We found that the compression modulus has very little influence on the maximum mass of pure nucleonic neutron stars: 
in the range $K=(200-300)$ MeV its influence is negligible for low effective masses $m_N^*/m_N < 0.6$, while 
its most prominent effect is for large effective masses $m_N^*/m_N > 0.7$ where it causes a mass difference of 
just $\Delta_M <$ 0.1${M}_{\odot}$.  
The effective mass of the nucleon at saturation proved to be a very good parameter for obtaining large maximum masses. 
Over the investigated range $m_N^*/m_N=0.55-0.8$ the maximum masses change about $\Delta_M \approx$ 1.1${M}_{\odot}$. 
The most massive stars (up to $\sim$2.9${M}_{\odot}$) are obtained for low effective nucleon masses independently of the compression modulus, 
while for $m_N^*/m_N > 0.78$ we cannot reach the mass limit of 1.97$\pm$0.04${M}_{\odot}$ if the compression modulus is too low. 
For the hyperonic case, maximum masses varied from 2.0-2.1${M}_{\odot}$ 
at $m_N^*/m_N=0.55$ to about 1.39-1.63${M}_{\odot}$ at $m_N^*/m_N=0.8$ in the ``model $\sigma\omega\rho$'', whereas
in the ``model $\sigma\omega\rho\phi$'' the variation is from 2.4 to 1.6 $M_{\odot}$ respectively.
Hyperonic neutron stars in the ``model $\sigma\omega\rho$'' are only compatible with
the new pulsar mass constraint for low values of the effective nucleon mass $m_N^*/m_N \leq 0.6$ MeV, while 
within the  ``model $\sigma\omega\rho \phi$'' the mass constraint can only be achieved for $m_N^*/m_N \leq 0.7$ MeV.
We conclude that the nuclear compression modulus at saturation might be a good indicator for the stiffness of the nuclear matter EoS at low densities (around saturation), 
but the maximum mass is much more sensitive to the effective nucleon mass at saturation. This result is in accordance
with the fact that the high density behaviour of the EoS in the RMF model is solely controlled by the effective nucleon 
mass at saturation density by virtue of the Hugenholtz-van Hove theorem \cite{Waldhauser88}.  \\
\indent  As we have seen, a 2${M}_{\odot}$ star like PSR J1614-2230 helps to constrain the nuclear matter EoS.
Then a possible 2.4${M}_{\odot}$ star - as has been recently suggested to exist \cite{vanKerkwijk} - would pose even tighter constraints on the neutron star composition. 
Should this measurement be confirmed, we could rule out many hadronic models (e.g. TM1 or GM3) even in the limit of purely nucleonic stars. 
Furthermore, in most such models the appearance of hyperons together with the use of 
SU(6) relations for the vector coupling constants would then become impossible. In an associated paper \cite{SimonDebbie2}, we further investigate the possibility of existence of hyperons
in massive neutron stars, by questioning the assumption of SU(6) symmetry for the determination of vector meson-hyperon couplings. \\
$\:$\\
\textit{Acknowledgements:} 
J.S.-B. is supported by the DFG through the Heidelberg Graduate School of Fundamental Physics. 
D.C. acknowledges the support from the Alexander von Humboldt foundation. 
S.W. is supported by the state of Baden-W\"urttemberg through a LGFG stipend. 
This work is supported by BMBF under grant FKZ 06HD9127, by the Helmholtz Alliance HA216/EMMI and by 
CompStar, a research networking program of the European Science Foundation.

\providecommand{\href}[2]{#2}\begingroup\raggedright
\endgroup
\end{document}